\documentclass[prl,aps,twocolumn]{revtex4}

\usepackage{graphicx,epsfig}
\usepackage{float}
\usepackage{dcolumn}
\usepackage{bm}

\begin{document}


\title{Correlation-Induced Resonances and Population Switching in a Quantum Dot Coulomb Valley}

\author{Hyun-Woo Lee}
\author{Sejoong Kim}
\altaffiliation[Present address: ]{Department of Physics,
Massachusetts Institute of Technology, Cambridge, MA 02139, USA}

\affiliation{PCTP and Department of Physics, Pohang University of Science
and Technology, Pohang, Kyungbuk 790-784, Korea
}

\date{\today}

\begin{abstract}
Strong correlation effects on electron transport through a spinless quantum dot are considered.
When two single-particle levels in the quantum dot are degenerate,
a conserved pseudospin degree of freedom appears
for generic tunneling matrix elements between the quantum dot and leads.
Local fluctuations of the pseudospin in the quantum dot give rise to
a pair of asymmetric conductance peaks near the center of a Coulomb valley.
An exact relation to the population switching is provided.
\end{abstract}


\maketitle

{\it Introduction.}---
Electronic transport through a quantum dot (QD) is a useful probe
of the strong Coulomb interaction effects in zero dimensional systems~\cite{Sohn97Book}.
One of well-known interaction effects is the Coulomb blockade~\cite{Kastner93PT},
which allows the current to  flow through the QD
only at special gate voltages
and suppresses the current at other gate voltages (Coulomb valleys).
The interaction also induces electronic correlations
responsible for deviations from the orthodox Coulomb blockade theory~\cite{Kastner93PT}.
In Coulomb valleys,
the current may be enhanced by the spin fluctuations
via the spin Kondo effect~\cite{SpinKondoTh,SpinKondoExp},
or by the orbital fluctuations
via the orbital Kondo effect~\cite{OrbitalKondoTh,Boese01PRB,OrbitalKondoExp}.

\begin{figure}[bbbp]
\begin{center}
\psfig{file=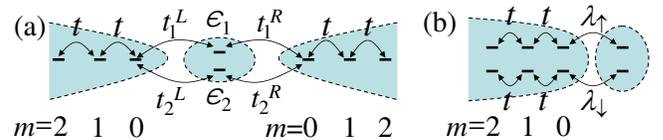,width=\columnwidth}
\caption{\label{modeling}
(Color online) (a) A quantum dot with two single-particle levels $\epsilon_1$, $\epsilon_2$
coupled to two leads. When $\epsilon_1=\epsilon_2$,
the system may be transformed
to a new system shown in (b).
}
\end{center}
\end{figure}

Recently an intriguing experimental report~\cite{Schuster97Nature} of the  anomalous transmission phase
through a QD motivated theoretical studies~\cite{Silvestrov00PRL,Sindel05PRB,Kim06PRB,Meden06PRL}
of the correlation effects in a spinless QD system with two single-particle levels [Fig.~\ref{modeling}(a)].
In particular, the study~\cite{Meden06PRL} using the functional renormalization group method
revealed that when the two levels are degenerate,
the conductance through the QD is anomalously enhanced near the center of a Coulomb valley
and forms a pair of asymmetric peaks.
These peaks are termed as correlation-induced resonances (CIRs).
The nature of the correlation, however, remains unclear.
The spin Kondo effect~\cite{SpinKondoTh,SpinKondoExp} is not applicable
since the system is spinless.
The orbital Kondo effect in Refs.~\cite{OrbitalKondoTh,Boese01PRB,OrbitalKondoExp}
is not applicable either
since it occurs only when the tunneling matrix elements between the QD and leads satisfy certain constraints~\cite{OrbitalKondoTh,Boese01PRB}
while the CIRs appear for generic tunneling matrix elements.
A possibly related phenomenon is
the so-called population switching (PS)~\cite{Silvestrov00PRL,Sindel05PRB,Kim06PRB};
Near the center of the Coulomb valley,
the electron population of the QD switches from one single-particle level to the other.
The relation between the CIRs and the PS also remains unclear however.
%
In this Letter,
(i) we show that the QD system with two single-particle levels
possesses  a conserved pseudospin degree of freedom
when the two levels are degenerate,
(ii) provide an exact relation between the CIRs and the PS,
and
(iii) demonstrate that local fluctuations of the pseudospin in the QD
are the origin of the CIRs.

%
The spinless QD system may be realized in experiments, for
instance, when a QD with two orbital levels, each with the
two-fold spin degeneracy, is subjected to a strong magnetic field.
If the resulting Zeeman splitting is sufficiently larger than
the energy difference of the two orbital levels,
the transport
through the QD in the Coulomb valley with only one electron in the
QD can be described by the following spinless
Hamiltonian~\cite{Sindel05PRB,Meden06PRL},
\begin{equation}\label{hamiltonian}
{\cal H}={\cal H}_{\rm dot}+{\cal H}_{\rm lead}+{\cal H}_{T},
\end{equation}
where
${\cal H}_{\rm dot}  \equiv
\sum_{j=1}^{2}\epsilon_j d_j^\dag d_j
+ U ( n_1 -1/2)( n_2 -1/2)$,
${\cal H}_{\rm lead} \equiv
-t \sum_{l=L,R}\sum^{\infty}_{m=0}
( c^\dag_{m,l}c_{m+1,l} +H.c. )$,
${\cal H}_{T} \equiv  -\sum_{j,l} ( t_j^{l} c_{0,l}^\dag d_j + H.c.)$.
Here $\epsilon_{j}$ denotes the  energy of the single-particle state $j$ in the QD.
$d_{j}$ and $c_{m,l}$ are the annihilation operators
for the electron in the QD
and for the electron at the site $m$ in the lead $l$,
respectively. $n_j\equiv d^\dagger_j d_j$. Note that each lead
contains only one channel [Fig.~\ref{modeling}(a)].
This is motivated by the experimental situation in Ref.~\cite{Schuster97Nature},
where narrow constrictions are introduced between a QD and
leads in order to force the system into the single channel regime.

{\it Pseudospin}.---
The pseudospin degree of freedom can be revealed by the following unitary transformations
from $c_{m,L(R)}$, $d_{1(2)}$ to the new operators $c_{m,\uparrow(\downarrow)}$, $d_{\uparrow(\downarrow)}$,
\begin{eqnarray}
\label{U_lead}
\left( \begin{array}{cc} c_{m,\uparrow} \\ c_{m,\downarrow}
\end{array} \right)
\!=\! {\cal U}_{\rm lead} \! \left( \! \begin{array}{cc} c_{m,L} \\ c_{m,R}
\end{array} \! \right),
& &
{\cal U}_{\rm lead}=\left(\begin{array}{cc} \eta & \zeta^{*}
\\ \zeta & -\eta^{*} \end{array}\right),
\\
\label{U_dot}
\left( \begin{array}{cc} d_{\uparrow} \\ d_{\downarrow}
\end{array} \right)
={\cal U}_{\rm dot} \left( \begin{array}{cc} d_{1} \\ d_{2}
\end{array} \right),
& &
{\cal U}_{\rm dot}=\left(\begin{array}{cc} \alpha & \beta^{*}
\\ \beta & -\alpha^{*} \end{array}\right),
\end{eqnarray}
under which ${\cal H}_{\rm lead}$ remains invariant
for general $\alpha$, $\beta$, $\eta$, $\zeta$
with $\left|\alpha\right|^2+\left|\beta\right|^2=\left|\eta\right|^2+\left|\zeta\right|^2=1$.
When the two dot levels are degenerate $\epsilon_1=\epsilon_2=\epsilon$,
${\cal H}_{\rm dot}$ transforms to
${\cal H}_{\rm dot}^{\rm (1)}\equiv
 \sum_{\sigma=\uparrow,\downarrow}\epsilon d_{\sigma}^{\dagger}d_{\sigma}
+U(n_\uparrow -{1 \over 2})( n_\downarrow - {1\over 2} )$,
and thus ${\cal H}_{\rm dot}$ also remains invariant.
Finally ${\cal H}_{T}$ transforms to
\begin{equation}
{\cal H}_{T}  =  -\left( \begin{array}{cc}
c_{0,\uparrow}^{\dagger} & c_{0,\downarrow}^{\dagger}\end{array} \right)
\Lambda
\left( \begin{array}{cc} d_{\uparrow} \\
d_{\downarrow}\end{array} \right) +H.c.,
\label{H-T-Lambda}
\end{equation}
where $\Lambda\equiv {\cal U}_{\rm lead}T {\cal U}_{\rm dot}^{\dagger}$
and $T=\left( \begin{array}{cc} t_1^L & t_2^L \\ t_1^R & t_2^R
\end{array} \right)$.
We choose ${\cal U}_{\rm lead}$ and ${\cal U}_{\rm dot}$
in such a way that $\Lambda$ becomes diagonal
with diagonal elements $\lambda_\uparrow$ and $\lambda_\downarrow$,
{\it i.e.},
${\cal H}_{T}=-( \lambda_{\uparrow}c_{0,\uparrow}^\dag d_\uparrow + \lambda_{\downarrow} c_{0,\downarrow}^\dag d_\downarrow + H.c.)$.
Such a diagonalization can be achieved for {\it general} $T$
by choosing ${\cal U}_{\rm lead}$ and ${\cal U}_{\rm dot}$
to be the solutions of the eigenvalue equations
$(TT^\dagger) {\cal U}_{\rm lead}^\dagger={\cal U}_{\rm lead}^\dagger(\Lambda^\dagger \Lambda)$
and $(T^\dagger T) {\cal U}_{\rm dot}^\dagger
={\cal U}_{\rm dot}^\dagger (\Lambda \Lambda^\dagger)$.
Without loss of generality, we may assume that both $\lambda_\uparrow$ and $\lambda_\downarrow$ are real
and $\lambda_\uparrow \ge \lambda_\downarrow \ge 0$.
Note that in the transformed system,
the electron tunneling between the pseudospin $\uparrow$ states [upper half in Fig.~\ref{modeling}(b)]
and the pseudospin $\downarrow$ states (lower half) is prohibited.
%
This illustrates the existence of the conserved pseudospin
$\sigma$ ($\uparrow$ or $\downarrow$)
for general $t_j^l$'s.
This generalizes the earlier reports~\cite{Boese01PRB,Meden06PRL}
of the conserved pseudospin for special $t_j^l$'s,
for which $\lambda_\uparrow=\lambda_\downarrow$
and the system possesses the SU(2) pseudospin symmetry.
In contrast, the symmetry is reduced to U(1) when $\lambda_\uparrow \ne \lambda_\downarrow$.
Later in this Letter, it will be demonstrated that
the difference $\lambda_\uparrow \neq \lambda_\uparrow$
is crucial for the CIRs.

On the other hand, when the degeneracy is lifted
$\epsilon_{1(2)}=\epsilon \pm \delta/2$,
the original Hamiltonian [Eq.~(\ref{hamiltonian})] transforms to
${\cal H}_{\rm dot}^{(1)}+{\cal H}_{\rm dot}^{(2)}+{\cal H}_{\rm dot}^{(3)}+{\cal H}_{\rm lead}+{\cal H}_T$
under the transformations [Eqs.~(\ref{U_lead}),(\ref{U_dot})] that
diagonalize $\Lambda$. Here the two additional terms ${\cal
H}_{\rm dot}^{(2)}$ and ${\cal H}_{\rm dot}^{(3)}$ are defined as
${\cal H}_{\rm dot}^{(2)} \equiv
\delta(|\alpha|^2-|\beta|^2)(n_\uparrow-n_\downarrow)/2$,
and ${\cal H}_{\rm dot}^{(3)} \equiv \delta(\alpha \beta^*
d_{\uparrow}^{\dagger}d_{\downarrow} +\beta \alpha^*
d_{\downarrow}^{\dagger}d_{\uparrow})$. Since
$(n_\uparrow-n_\downarrow)/2$ amounts to the QD pseudospin along
the pseudospin quantization axis, say $z$, ${\cal H}_{\rm
dot}^{(2)}$ can be interpreted as the Zeeman coupling to the
parallel
pseudo-magentic field $H^\delta_z=-\delta (|\alpha|^2-|\beta|^2)$
along the $z$-axis.
${\cal H}_{\rm dot}^{(2)}$ preserves the pseudospin conservation.
On the other hand, ${\cal H}_{\rm dot}^{(3)}$ can be interpreted
as the Zeeman coupling to the perpendicular pseudomagnetic field,
whose $x$-component is given by $H^\delta_x=-2\delta{\rm
Re}(\alpha \beta^*)$ and $y$-component by $H^\delta_y=2\delta{\rm
Im}(\alpha \beta^*)$. ${\cal H}_{\rm dot}^{(3)}$ breaks the
pseudospin conservation along the $z$-axis.

\begin{figure}[ttbp]
\begin{center}
\psfig{file=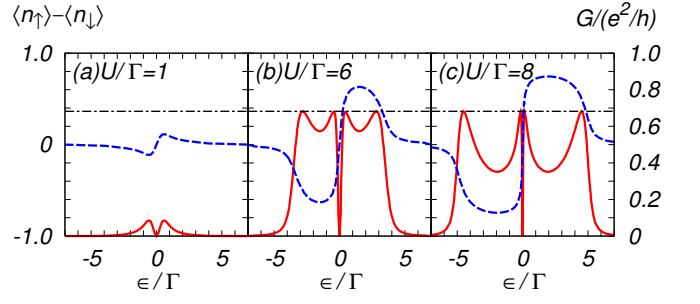,width=\columnwidth}
\caption{
\label{conductance_population}
(Color online) The relation between the conductance $G$ (red solid line) and the population difference
$\langle n_\uparrow \rangle-\langle n_\downarrow \rangle$ (blue dashed line)
when the two dot levels are degenerate $\epsilon_1=\epsilon_2=\epsilon$.
%
In all three panels, $t_1^L:t_1^R:t_2^L:t_2^R=\sqrt{0.27}:\sqrt{0.33}:\sqrt{0.16}:-\sqrt{0.24}$.
The black horizontal dash-dotted line represents $G_{\rm max}\approx 0.68 (e^2/h)$ predicted by Eq.~(\ref{general_conductance_1}).
Here $\Gamma\equiv \pi[(|t_1^L|^2+|t_2^L|^2)\rho^L+(|t_1^R|^2+|t_2^R|^2)\rho^R]$, where
$\rho^l$ is the local density of states at the end of the lead $l$.
The curves for $G$ are from Fig.~2 in Ref.~\cite{Meden06PRL}
while the curves for $\langle n_\uparrow \rangle-\langle n_\downarrow \rangle$ are obtained~\cite{comment-fig2}
from Eq.~(\ref{general_conductance_1}).
%
(a) For a small $U$, where the PS (sign change of $\langle n_\uparrow \rangle - \langle n_\downarrow \rangle$ at $\epsilon=0$) is a weak feature
and $|\langle n_\uparrow \rangle - \langle n_\downarrow \rangle|$ remains below the critical value 1/2,
only two conductance peaks appear with the peak centers located at the positions
where $|\langle n_\uparrow \rangle - \langle n_\downarrow \rangle|$ is maximized.
(b) For a large $U$, where the PS becomes manifest and
$|\langle n_\uparrow \rangle - \langle n_\downarrow \rangle|$ becomes larger than
the critical value 1/2 in certain ranges of $\epsilon$,
four conductance peaks appear with the peak centers located at the positions where
$|\langle n_\uparrow \rangle - \langle n_\downarrow \rangle|=1/2$;
the two peaks close to $\epsilon=0$ are the CIRs
while the other two peaks are the Coulomb blockade peaks.
(c) For a still larger $U$, where the PS becomes stronger,
the distinction between the CIRs and the Coulomb blockade peaks becomes more evident.
Note that $G$ is related to the population difference
of the transformed dot states,
$\langle n_\uparrow \rangle - \langle n_\downarrow \rangle$,
instead of the population difference of the original dot states,
$\langle n_1 \rangle - \langle n_2 \rangle$.
%
}
\end{center}
\end{figure}

{\it CIRs vs. PS}.---
To examine transport properties for the degenerate case,
we first construct
the $\uparrow$- and $\downarrow$-scattering states in the transformed system [Fig.~\ref{modeling}(b)],
$\psi_{\uparrow}(x) = \chi_{\uparrow}
(e^{+ikx} + e^{2i{\theta}_\uparrow}e^{-ikx})$,
$\psi_{\downarrow}(x) = \chi_\downarrow
(e^{+ikx} + e^{2i{\theta}_\downarrow}e^{-ikx})$,
where $\chi_\uparrow$ and $\chi_\downarrow$ denote the spinors representing
the pseudospin $\uparrow$ and $\downarrow$ states, respectively,
and $x \propto -m (\le 0)$.
Note that the pseudospin flip between $\uparrow$ and $\downarrow$ states
is prohibited in the scattering states due to the pseudospin conservation.
From the Friedel sum rule~\cite{Langer61PR},
the scattering phases $\theta_{\uparrow}= \pi\langle n_\uparrow \rangle$
and $\theta_{\downarrow} = \pi\langle n_\downarrow \rangle$,
where $\langle n_\sigma \rangle$ denotes the expectation value of $n_\sigma\equiv d_\sigma^\dagger d_\sigma$
with respect to the ground state.
Next we take proper coherent superpositions (see for instance Ref.~\cite{Lee99PRL}) 
of $\psi_{\uparrow}$ and $\psi_{\downarrow}$
to evaluate the transmission probability
in the {\it original} system [Fig.~\ref{modeling}(a)].
Then from the Landauer-B\"uttiker formula, one obtains
the zero temperature conductance $G$,
\begin{equation}\label{general_conductance_1}
G=G_{\rm max}
\sin^2\left[\pi\left(\langle n_\uparrow \rangle - \langle n_\downarrow
\rangle\right)\right],
\end{equation}
where
$G_{\rm max}/(e^2/h)=
4|t_1^L{t_1^R}^*+t_2^L{t_2^R}^*|^2/
[(|t_1^L|^2+|t_2^L|^2-|t_1^R|^2-|t_2^R|^2)^2
+4|t_1^L{t_1^R}^*+t_2^L{t_2^R}^*|^2]$.
As illustrated in Fig.~\ref{conductance_population},
Eq.~(\ref{general_conductance_1}) provides a relation between the CIRs and the PS.

It is illustrative to compare Eq.~(\ref{general_conductance_1})
with the corresponding expression
for the conventional spin Kondo effect in a QD~\cite{SpinKondoTh},
where spin-up and spin-down scattering states generates
two {\it incoherent} contributions
[$\sin^2(\pi \langle n_\uparrow \rangle)$ vs. $\sin^2(\pi \langle n_\downarrow\rangle )]$
to the conductance.
Thus in the absence of a magnetic field, where $\langle n_\uparrow \rangle=\langle n_\downarrow \rangle$,
$G$ is proportional to $\sin^2(\pi \langle n_\uparrow \rangle)+\sin^2(\pi \langle n_\downarrow \rangle)
=2\sin^2[\pi ( \langle n_\uparrow \rangle+\langle n_\downarrow \rangle)/2]$.
In our system, in contrast,
the two pseudospin scattering states $\psi_\uparrow$, $\psi_\downarrow$
should be {\it coherently} superposed to construct
scattering states
in the original system,
which are then used to evaluate $G$.
This coherent summation procedure
takes into account
the interference between the two transport paths in Fig.~\ref{modeling}(a),
one mediated by $\epsilon_1$ and the other by $\epsilon_2$.
This explains the difference between the two expressions for $G$.

{\it Pseudospin fluctuations}.---
To examine fluctuations of the QD pseudospin $S_z\equiv (n_\uparrow-n_\downarrow)/2$,
it is useful to map the Hamiltonian ${\cal H}$
into a $s$-$d$ model
by using the Schrieffer-Wolff transformation~\cite{Hewson93Book}.
In the large $U$ limit and near the center of the Coulomb valley,
one finds~\cite{Kdiagonalization},
\begin{equation}
{\cal H}_{s-d}={\cal H}_{\rm ex}+{\cal H}_{\rm lead}
+{\cal H}_{B},
\label{H_sd}
\end{equation}
where
${\cal H}_{\rm ex} = \sum_{kk'} [J^+ S^+
c^\dagger_{k\downarrow}c_{k'\uparrow}
+J^-S^- c^\dagger_{k\uparrow}c_{k'\downarrow}
+  J^z S_z (c^\dagger_{k\uparrow}c_{k'\uparrow}-c^\dagger_{k\downarrow}c_{k'\downarrow})]$,
${\cal H}_B = -S_z {\cal B}_z^{\rm eff}$,
and
${\cal B}_z^{\rm eff} = B_z-\sum_{kk'}J^{z,2}_{kk'}(c^\dagger_{k\uparrow}c_{k'\uparrow}+c^\dagger_{k\downarrow}c_{k'\downarrow})$.
Here $c_{k\sigma}$ is the annihilation operator of the eigenstate
with energy $\varepsilon_k$ in ${\cal H}_{\rm lead}$.
The degeneracy of the dot states is still assumed.
Various coefficients are defined as follows;
$J^+=J^-=4 V_\uparrow V_\downarrow /U$,
$J^z=2 (V_\uparrow^2+V_\downarrow^2)/U$,
%
$B_z= (V_{\uparrow}^2 - V_{\downarrow}^2)\sum_k (U/2-\epsilon+\varepsilon_k)^{-1}$,
and $J^{z,2}_{kk'}=(V_{\uparrow}^2-V_{\downarrow}^2) [(U/2+\epsilon-\varepsilon_{k'})^{-1} +(U/2-\epsilon+\varepsilon_k)^{-1}]/2$,
where
$V_{\sigma}\ge 0$
($V_{\uparrow}/V_{\downarrow}=\lambda_{\uparrow}/\lambda_{\downarrow}$) denotes
the matrix element for the tunneling from the dot state $d_{\sigma}$
to the lead state $c_{k\sigma}$.
Note that ${\cal H}_{\rm ex}$ becomes an {\it anisotropic} antiferromagnetic ($J^z >J^+> 0$)
exchange interaction
since $V_\uparrow \neq V_\downarrow$ in general.
%
A crucial difference from the conventional Kondo effects~\cite{Hewson93Book} arises from
the pseudomagnetic field ${\cal B}_z^{\rm eff}$,
whose expectation value $H_z$ with respect to the Fermi sea in the leads becomes~\cite{comment-pseudo-B}
\begin{equation}\label{pseudo_B}
H_z\equiv \langle {\cal B}_z^{\rm eff}\rangle
={\Gamma_\uparrow-\Gamma_\downarrow \over \pi}\ln {U/2+\epsilon \over
U/2-\epsilon}\, ,
\end{equation}
where $\Gamma_\sigma\equiv \pi \rho_0 V_\sigma^2$
and $\rho_0$ is the density of states in the leads.
Note that $H_z$ does depend on $\epsilon$ and changes its sign at $\epsilon=0$,
implying the sign change of $\langle S_z \rangle=(\langle n_\uparrow \rangle
-\langle n_\downarrow \rangle)/2$ at $\epsilon=0$.
This provides a simple explanation of the PS~\cite{Silvestrov00PRL,Sindel05PRB,Kim06PRB}.
By the way, for the special $t_j^l$'s discussed in Refs.~\cite{Boese01PRB,Meden06PRL},
where the conserved pseudospin exists but $\lambda_\uparrow=\lambda_\downarrow$,
$H_z$ vanishes since $\Gamma_\uparrow=\Gamma_\downarrow$.

Next we perform
the two-stage poor man's scaling~\cite{Hewson93Book},
the first stage with the original Hamiltonian ${\cal H}$
and the second stage with ${\cal H}_{s-d}$
up to the second order in $J^z$ and $J^\pm$.
One finds that the $S_z$ fluctuations are characterized by
the Kondo temperature $T_K$~\cite{TK-comment},
\begin{equation}
\label{T-Kondo}
T_K \sim U \exp\left(-{1 \over 4\rho_0 J_0}\ln {J^z+J_0 \over
J^z-J_0} \right),
\end{equation}
where $J_0\equiv \sqrt{(J^z)^2-(J^+)^2}$.
Equation~(\ref{T-Kondo}) may be expressed as
$T_K \sim U\exp[-\pi U \ln (\Gamma_\uparrow/\Gamma_\downarrow)/ 8(\Gamma_\uparrow - \Gamma_\downarrow) ]$.
Interestingly ${\cal B}_z^{\rm eff}$ at each step of the scaling
shares the same expectation value $H_z$ [Eq.~(\ref{pseudo_B})].

\begin{figure}[ttbp]
\begin{center}
\psfig{file=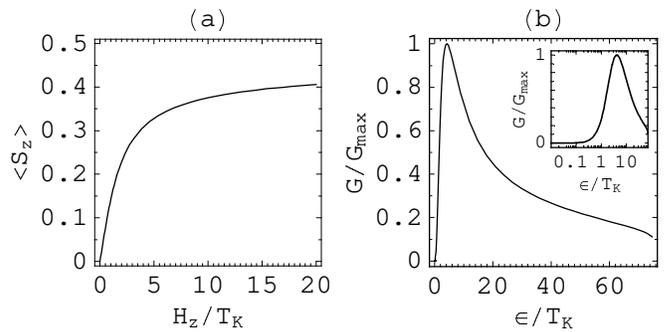,width=\columnwidth}
\caption{\label{magnetization}
(a) $\langle S_z \rangle=(\langle n_\uparrow \rangle - \langle n_\downarrow \rangle)/2$
as a function of $H_z/T_K$ predicted by the exact solution
of the anisotropic $s$-$d$ model~\cite{Tsvelick83AP}.
$\langle S_z \rangle$ for negative $H_z$ can be obtained
by using $\langle S_z \rangle$ being an odd function of $H_z$.
(b) $G/G_{\rm max}$ vs. $\epsilon/T_K$~\cite{TK-comment}
obtained by Eqs.~(\ref{general_conductance_1}), (\ref{pseudo_B}), and
the $\langle S_z \rangle$ vs. $H_z$ relation in (a).
In these plots,
$t_1^L:t_1^R:t_2^L:t_2^R=\sqrt{0.27}:\sqrt{0.33}:\sqrt{0.16}:-\sqrt{0.24}$
($\lambda_\uparrow/\lambda_\downarrow\approx 1.29$),
and $U/(\Gamma_\uparrow+\Gamma_\downarrow)=7$ are used.
$G/G_{\rm max}$ for negative $\epsilon$ can be obtained
by using $G$ being an even function of $\epsilon$.
In the inset, the logarithmic scale is used for the horizontal axis.
}
\end{center}
\end{figure}

For further study,
we approximate ${\cal H}_{s-d}$
by replacing ${\cal B}_z^{\rm eff}$ with $H_z$.
Properties of the resulting Hamiltonian are well known
via the Bethe ansatz method~\cite{Tsvelick83AP}.
Figure~\ref{magnetization}(a) shows the $H_z$-dependence of $\langle S_z \rangle$
predicted by the exact solution~\cite{Tsvelick83AP},
and Fig.~\ref{magnetization}(b) shows the resulting $\epsilon$-dependence of $G/G_{\rm max}$
obtained from Eq.~(\ref{general_conductance_1}).
For $|H_z| \lesssim T_K$, $\langle S_z \rangle$ is approximately given
by $H_z/(2\pi T_K)$~\cite{Tsvelick83AP}.
Within this linear approximation,
the CIR peak positions $\epsilon_{\rm CIR}$ are
given by
\begin{equation}
\label{V_CIR}
\epsilon_{\rm CIR}\approx \pm {U \over 2}\tanh \left(
{\pi^2 \over 2}{T_K \over \Gamma_\uparrow-\Gamma_\downarrow} \right).
\end{equation}
For $H_z \gg T_K$, $1/2-\langle S_z \rangle$ is proportional
to $(T_K/H_z)^{2\mu /\pi}$~\cite{Tsvelick83AP},
where $\mu\equiv \cos^{-1}[\cos(2\pi \rho_0 J^z)/\cos(2\pi \rho_0 J^+)]\approx
2\pi\rho_0 J_0$ in the weak tunneling regime $\rho_0 J^z$, $\rho_0 J^+ \ll 1$.
Due to the small exponent $2\mu/\pi\approx 4 \rho_0 J_0 \ll 1$,
$\langle S_z \rangle$ approaches its saturated value $1/2$ very slowly.
Combined with Eqs.~(\ref{general_conductance_1}) and (\ref{V_CIR}),
this explains the origin of the strongly asymmetric peak shape
of the CIRs~\cite{Meden06PRL}.

{\it Discussion.}---
First we address the case of the nondegenerate dot levels
$\epsilon_{1(2)}=\epsilon \pm \delta/2$.
For small $\delta$, the two perturbations ${\cal H}_{\rm dot}^{(2)}$, ${\cal H}_{\rm dot}^{(3)}$
due to the nondegeneracy may be treated separately.
Effects of ${\cal H}_{\rm dot}^{(2)}$ are rather trivial;
After the Schrieffer-Wolff transformation,
${\cal H}_{\rm dot}^{(2)}$ becomes $-S_z H_z^\delta$,
which renormalizes $H_z$ in Eq.~(\ref{pseudo_B})
and induces
a shift of the CIR peaks to the new positions
$\epsilon_{\rm CIR} \approx \pm (U/2)
\tanh[\pi^2(T_K\mp 2H_z^\delta/\pi)/2(\Gamma_\uparrow-\Gamma_\downarrow)]$.
${\cal H}_{\rm dot}^{(2)}$ does not alter the peak heights.
Effects of ${\cal H}_{\rm dot}^{(3)}$ are rather complicated
since it breaks the pseudospin conservation.
Equation~(\ref{general_conductance_1}) based on the pseudospin conservation
is not applicable
and we derive below a more general conductance formula.
The time-reversal symmetry is assumed for simplicity.
Electrons in the transformed lead [Fig.~\ref{modeling}(b)] can be described by
the following scattering matrix,
\begin{equation}
\left(\begin{array}{cc} r_{\uparrow\uparrow} & \tau_{\uparrow\downarrow}
\\ \tau_{\downarrow\uparrow} & r_{\downarrow\downarrow} \end{array}\right)
=e^{2i\theta_{\rm tot}}
\left(\begin{array}{cc} e^{2i\theta}\cos\phi &
i\sin\phi
\\ i\sin\phi & e^{-2i\theta}\cos\phi \end{array}\right),
\end{equation}
where $\theta_{\rm tot}= \pi(\langle n_\uparrow \rangle +\langle n_\downarrow \rangle)/2=\pi/2$
in the Coulomb valley due to the Friedel sum rule~\cite{Langer61PR}.
After a similar algebra as in the derivation of Eq.~(\ref{general_conductance_1}), one obtains
\begin{equation}\label{transmission_with_nondegeneracy}
G=G_{\rm max}
|\sin2\theta\cos\phi-\cot2\nu\sin\phi|^2,
\end{equation}
where $\nu$ is independent of $\epsilon$ with
$\cot^2 2\nu=(|t_1^L|^2+|t_2^L|^2-|t_1^R|^2-|t_2^R|^2)^2/4|t_1^L {t_1^R}^* +t_2^L {t_2^R}^*|^2$
while $\theta$ and $\phi$ in general depend on $\epsilon$.
For small $\delta$, the $\epsilon$-dependence of $\theta$ and $\phi$ can
be estimated from the knowledge in the limit $\delta\rightarrow 0$,
where the pseudospin flip amplitudes
$\tau_{\uparrow\downarrow}=\tau_{\downarrow\uparrow}$ approach zero
and thus $\phi$ and $\theta$ approach respectively to zero
and $\pi(\langle n_\uparrow \rangle-\langle n_\downarrow \rangle)/2$.
Then in generic situations,
where $\tau_{\uparrow\downarrow}=\tau_{\downarrow\uparrow}$
do not vanish in the narrow range of $\epsilon$ near the two CIRs,
the second term in Eq.~(\ref{transmission_with_nondegeneracy})
does not change its sign near the CIRs
whereas the first term changes its sign due to the sign reversal of $\sin 2\theta$.
This implies that
while the interference between the two terms is destructive near one CIR peak,
suppressing the peak height,
it is constructive near the other CIR peak,
enhancing the peak height.
This explains the $\delta$-induced difference of
the two CIR peak heights reported in Ref.~\cite{Meden06PRL}.

Next we remark briefly on the conductance at the dip, $G_{\rm dip}$, between the two CIR peaks.
For $\delta=0$, the PS always results in $G_{\rm dip}=0$ [Eq.~(\ref{general_conductance_1})].
For nonzero but small $\delta$,
$G_{\rm dip}$ should be still exactly $0$
if the system has the time-reversal symmetry
since the exact cancellation of the two terms in Eq.~(\ref{transmission_with_nondegeneracy}) is possible.
If the time-reversal symmetry is broken, a further generalization of
Eq.~(\ref{transmission_with_nondegeneracy}) indicates that
such an exact cancellation is not generic and $G_{\rm dip}$
acquires a finite value.
This result is consistent with Ref.~\cite{Lee99PRL,Silvestrov03PRL}.


In summary, we have demonstrated that
a spinless quantum dot system with two degenerate single-particle levels
allows a conserved pseudospin
and that
in the presence of the correlation caused by the strong Coulomb interaction,
the fluctuations of the pseudospin at the quantum dot give rise to a pair
of asymmetric conductance peaks in a Coulomb valley.
The relation between these correlation-induced resonances
and the phenomenon of the population switching has been established.

This work was supported by the SRC/ERC program
(R11-2000-071) and the Basis Research Program (R01-2005-000-10352-0) of MOST/KOSEF,
by the POSTECH Core Research Program, and by the KRF Grant (KRF-2005-070-C00055)
funded by MOEHRD.

{\it Note added.}--- After the submission of our paper,
preprints~\cite{Silvestrov06cond-mat,Kashcheyevs06cond-mat}
reporting similar results appeared.


\begin{thebibliography}{99}
\bibitem{comment-fig2} Since the mapping from $G$ to $\langle n_\uparrow \rangle-\langle n_\downarrow \rangle$
in Eq.~(\ref{general_conductance_1}) is multi-valued,
we impose the following constraints
to resolve the ambiguity;
(i) both $\langle n_\uparrow \rangle$ and $\langle n_\downarrow \rangle$ approach 0
as $\epsilon  \rightarrow +\infty$
and 1 as $\epsilon  \rightarrow -\infty$,
(ii) $\langle n_\uparrow \rangle-\langle n_\downarrow \rangle$ is positive for $\epsilon >0$
and negative for $\epsilon <0$ (due to the PS mechanism in Ref.~\cite{Silvestrov00PRL}),
(iii) $d(\langle n_\uparrow\rangle-\langle n_\downarrow \rangle )/d\epsilon $ is a continuous function of $\epsilon$,
(iv) near the conductance peaks, $G-G_{\rm max} \propto (\epsilon-\epsilon_{\rm peak})^2$,
where $\epsilon_{\rm peak}$ denotes the peak center position.
\bibitem{Kdiagonalization}
In addition to the terms in Eq.~(\ref{H_sd}),
the Schrieffer-Wolff transformation also generates
a potential scattering term.
This term is not essential and thus ignored.
\bibitem{TK-comment} The prefactor of $T_K$ in Eq.~(\ref{T-Kondo}) is not certain.
For example, higher order poor man's scaling may modify the prefactor.
In Fig.~\ref{magnetization}(b),
the prefactor is assumed to be exactly $U$ for definiteness of the illustration.
\end{thebibliography}


\end{document}